\documentclass[%
 reprint,
 amsmath,amssymb,
 aps,
]{revtex4-1}

\usepackage{graphicx}
\usepackage{dcolumn}
\usepackage{bm}
\usepackage{hyperref}

\usepackage[T1]{fontenc} 
\usepackage{xcolor}

\usepackage{enumitem}

\newcommand{\bei}{\begin{itemize}}
\newcommand{\eei}{\end{itemize}}
\newcommand{\bee}{\begin{enumerate}}
\newcommand{\eee}{\end{enumerate}}
\newcommand{\beeL}{\begin{enumerate}[label=(\Alph*)]}
\newcommand{\beel}{\begin{enumerate}[label=(\alph*)]}
\newcommand{\beeR}{\begin{enumerate}[label=(\Roman*)]}
\newcommand{\beer}{\begin{enumerate}[label=(\roman*)]}
\newcommand{\beeLd}{\begin{enumerate}[label=\Alph*.]}
\newcommand{\beeld}{\begin{enumerate}[label=\alph*.]}
\newcommand{\beeRd}{\begin{enumerate}[label=\Roman*.]}
\newcommand{\beerd}{\begin{enumerate}[label=\roman*.]}

\newcommand{\phase}{\theta}

\newcommand{\ov}{\over}
\newcommand{\g}{\gamma}

\newcommand{\xbr}{\xi}
\newcommand{\ubr}{\nu}

\newcommand{\adsst}{$\text{AdS}_3\times \text{S}^3\times \text{T}^4$ }

\def\bZ{{\mathbb Z}}
\newcommand{\nn}{\nonumber}
\newcommand{\la}{\label}

\def\bes{{\text{\tiny BES}}}
\def\afs{{\text{\tiny AFS}}}
\def\hl{{\text{\tiny HL}}}
\def\barnes{{\text{\tiny odd}}}

\renewcommand{\L}{{\scriptscriptstyle\text{L}}}
\newcommand{\R}{{\scriptscriptstyle\text{R}}}

\def\ka {{\kappa}}

\def\pa {\partial}

\def\cO{{\cal O}}

\def\cR{{\cal R}}

\newcommand{\tx}{\tilde{x}}
\newcommand{\tPhi}{\tilde{\Phi}}
\newcommand{\tg}{\tilde{\gamma}}

\begin{document}


\title{Dressing Factors for Mixed-Flux  \texorpdfstring{$AdS_3\times S^3\times T^4$}{AdS3xS3xT4} Superstrings}

\author{ Sergey Frolov}
 \email{frolovs@maths.tcd.ie}
 \affiliation{Hamilton Mathematics Institute and School of Mathematics Trinity College, Dublin 2, Ireland,}
 \affiliation{Istituto Nazionale di Fisica Nucleare, Sezione di Padova, via Marzolo 8, 35131 Padova,
Italy.}
\author{Davide Polvara}
 \email{davide.polvara@desy.de}
 \affiliation{%
 Dipartimento di Fisica e Astronomia, Universit\`a degli Studi di Padova, via Marzolo 8,
35131 Padova, Italy,
}%
\affiliation{Istituto Nazionale di Fisica Nucleare, Sezione di Padova, via Marzolo 8, 35131 Padova,
Italy,}%
\affiliation{II.~Institut f\"ur Theoretische Physik, Universit\"at Hamburg,
Luruper Ch.~149, 22761 Hamburg, Germany}
\author{Alessandro Sfondrini}
 \email{alessandro.sfondrini@unipd.it}
 \altaffiliation{MATRIX Simons Fellow.}
 \affiliation{%
 Dipartimento di Fisica e Astronomia, Universit\`a degli Studi di Padova, via Marzolo 8,
35131 Padova, Italy,
}%
\affiliation{Istituto Nazionale di Fisica Nucleare, Sezione di Padova, via Marzolo 8, 35131 Padova,
Italy.}
\date{\today}

\begin{abstract}
\noindent
We propose the dressing factors for the scattering of massive particles on the worldsheet of mixed-flux  $AdS_3\times S^3\times T^4$ superstrings, in the string and mirror kinematics. The proposal passes all self-consistency checks in the both kinematics, including for bound states. It matches with perturbative and semiclassical computations from the string sigma model, and with its relativistic limit.
\end{abstract}

\maketitle



Superstring theory  on  \adsst  backgrounds supported by a mixture of Ramond-Ramond (RR) and Neveu-Schwarz-Neveu-Schwarz (NSNS) fluxes is an interesting model playing important role  in the context of the
AdS/CFT correspondence \cite{Maldacena:1997re}. Similarly to what happens for $\text{AdS}_5$ and $\text{AdS}_4$ strings, see~\cite{Arutyunov:2009ga,Beisert:2010jr}, 
the $\text{AdS}_3$  model  is believed to be integrable if the string is quantised in a properly-chosen lightcone gauge, see \textit{e.g.}~\cite{Sfondrini:2014via}; This holds for any combination of the fluxes~\cite{Cagnazzo:2012se}.  In the lightcone gauge the model also possesses enough supersymmetry  to fix the worldsheet scattering matrix up to several functions, the so-called \textit{dressing factors} \cite{Hoare:2013ida,Lloyd:2014bsa}. They satisfy crossing equations which can be solved  under some assumptions on the analytic structure of the theory, see~\cite{Arutyunov:2004vx,Janik:2006dc,Beisert:2006ez} for the $\text{AdS}_5$ and $\text{AdS}_4$ strings. 

In the integrability approach, the knowledge of dressing factors is necessary to derive mirror Thermodynamic Bethe Ansatz (TBA) equations which determine the exact spectrum of the model  for any value of the tension and of the $B$-field. While for pure-NSNS backgrounds the spectrum can be found by worldsheet-CFT techniques~\cite{Maldacena:2000hw}, this is very hard in the presence of a finite amount of RR flux~\cite{Cho:2018nfn}, which leaves worldsheet integrability as the most viable road to solve mixed-flux models, see~\cite{Demulder:2023bux}.

For \adsst strings, only  the dressing factors for pure-RR~\cite{Frolov:2021fmj} and pure-NSNS~\cite{Baggio:2018gct} theories have been proposed. In the latter case, the mirror TBA reproduces the results of the worldsheet-CFT approach~\cite{Dei:2018mfl}, while in the former --- for pure-RR backgrounds, \textit{i.e.}~in absence of any $B$-field --- integrability yielded new quantitive predictions: The mirror TBA equations were derived in~\cite{Frolov:2021bwp} and used to study the tensionless limit of the spectrum~\cite{Brollo:2023pkl,Brollo:2023rgp} and the twisted ground-state energy~\cite{Frolov:2023wji}. For this case, a set of quantum spectral curve equations (which should be equivalent to the mirror TBA, at least in a subsector of the model related to~$\text{AdS}_3\times \text{S}^3$) had also been independently proposed~\cite{Ekhammar:2021pys,Cavaglia:2021eqr} and quantitatively investigated~\cite{Cavaglia:2022xld}.

Fixing the  dressing factors for the mixed-flux model  remains an open challenge since 2014~\cite{Lloyd:2014bsa} mainly due to the unique and intricate analytic structure of the model, though progress was recently made in unraveling this structure~\cite{Frolov:2023lwd} and in constructing the simplest pieces (the so-called \textit{odd parts}~\cite{Beisert:2006ib}) of the dressing factors~\cite{OhlssonSax:2023qrk}.

The peculiar features of this model can already be seen from  its non-relativistic dispersion relation~\cite{Hoare:2013lja}
\begin{equation}
\label{eq:dispersion}
    E(M,p)= \sqrt{\left(M+\frac{k}{2\pi}p\right)^2+4h^2\sin^2\frac{p}{2}}\,.
\end{equation}
Here $k=0,1,2,\dots$ and $h\geq0$ correspond to the strength of the NSNS and RR background fluxes, respectively. Particle multiplets of the model are distinguished by a $u(1)$ charge $M\in \bZ$ whose modulus $m\equiv |M|$ in the near-pp-wave expansion ~\cite{Berenstein:2002jq,Hoare:2013pma} (that is, at small~$p$ and large string tension) is proportional to the mass of  particles from a given multiplet; Hence, in what follows we loosely refer to~$m$ as the \textit{mass}.  All particles are then divided into three groups: i) \textit{massless} particles with $M=0$, ii) \textit{left} particles with $M=m>0$, iii) \textit{right} particles with $M=-m<0$. Massive left and right particles with $m=2,3,\dots$ are bound states of left and right particles of mass 1, respectively, but left and right particles do not form bound states. 
Taking into account the symmetries of the model, one finds that it is sufficient to know the dressing factors $\sigma_{\R\R}^{11}$, 
$\sigma_{\R\L}^{11}$, $\sigma_{\L}^{10}$
and $\sigma^{00}$ for the scattering of particles of mass $1$ and $0$, while all the other dressing factors are fixed by fusion, unitarity, and a ``left-right'' symmetry between left and right particles \cite{Lloyd:2014bsa}.

In this letter we propose a solution of the mixed-flux crossing equations for massive particles compatible with the symmetries of the model and with the available perturbative and semiclassical considerations~\cite{Hoare:2013pma,Berenstein:2002jq,Engelund:2013fja,Hoare:2013lja,Babichenko:2014yaa,Bianchi:2014rfa,Roiban:2014cia,Sundin:2014ema,Stepanchuk:2014kza}.

\paragraph{Worldsheet (``string'') kinematics.} 
\label{sec:crossing}

The kinematics of the worldsheet model is best expressed using $\ka$-deformed Zhukovsky variables 
\cite{Hoare:2013lja} which are functions of a  $u$-plane variable. They are
\begin{equation}
\label{uplane}
u_a(x)=x+\frac{1}{x}- \frac{\kappa_a}{\pi}\,\ln x  \quad\Leftrightarrow \quad 
x=\begin{cases}
x_a(u)\,,&\\
1/x_{\bar a}(u)\,,&      
\end{cases}
\end{equation}
with $x_a(u)^*=x_a(u^*)$, where $\ln x$ denotes the principal branch of~$\log x$.
Here and in what follows $a=\,$L,R.  If $a=\,$L then $\bar a=\,$R and vice versa; finally $\ka_\L=\ka=\tfrac{k}{h}$, $\ka_\R=-\ka$. 
We define the  $\ka$-deformed Zhukovsky variables for a particle of charge $M$ as usual
\begin{equation}\begin{aligned}
x_{a}^{\pm m}(u)=x_{a}(u\pm {im\ov h})\,,
\end{aligned}\end{equation} 
where $m=|M|$, and  we refer to  the variables 
\begin{equation}\begin{aligned}
p_a = i\,(\ln x_a^{-m} - \ln x_a^{+m})\,,
\end{aligned}\end{equation}
as momenta of the particles. Note that for real $u$ the range of the momenta is from 0 to $2\pi$. In terms of $x^{\pm m}_a$ the energy is
\begin{equation}
    E_a=\frac{h}{2i}\left(x^{+m}_a-\frac{1}{x^{+m}_a}-x^{-m}_a+\frac{1}{x^{-m}_a}\right).
\end{equation}
In the case $m=1$ we just write $x_{\L}^{\pm}(u)$, $x_{\R}^{\pm}(u)$.

\begin{figure}[t]
\includegraphics[width=4.2cm]{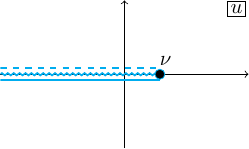}%
\hspace{2mm}%
\includegraphics[width=4.2cm]{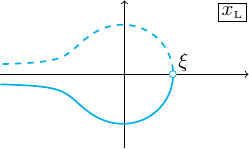}\\
\includegraphics[width=4.2cm]{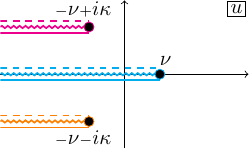}%
\hspace{2mm}%
\includegraphics[width=4.2cm]{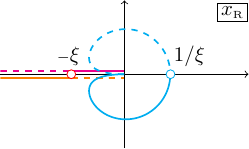}\\%
\caption{\label{fig:string}%
Analytic structure of $x_{\L,\R}(u)$. Top left: $x_{\L}(u)$ has a single branch cut (zigzag line). The images of its edges give an unbounded curve in the $x_{\L}$-plane (top right).
Bottom left: $x_{\R}(u)$ has two more branch cuts (magenta and orange). Their images are straight lines in the $x_{\R}$ plane, just above/below the cut of $\ln x$ (bottom right). The cyan curves split each $x_{\L,\R}$ planes in two parts, 
the one including $x_{\L,\R}=+\infty$
is the physical region, or ``string'' region; The one inside is the ``antistring'' region, where antiparticles live.}
\end{figure}

\paragraph{Analytic structure.}
The functions $x_\L(u)$ and $x_\R(u)$ have branch points, and we choose the cut structure described in~\cite{Frolov:2023lwd}: $x_\L(u)$ has one cut on its $u$-plane, for real $u<\ubr$, where $\ubr$ is the branch point; $x_\R(u)$ has the same branch cut, plus two more horizontal branch cuts stemming from $-\ubr\pm i\kappa$~\footnote{A very different cut structure of $x_a(u)$ was used in \cite{OhlssonSax:2023qrk}. Their identification of the physical regions also dramatically differs from ours. Their choice seems to be inconvenient for the analytic continuation to the mirror theory.}, see Figure~\ref{fig:string}.
The images of $\ubr$ on the $x_{\L,\R}$-planes sit at
\begin{equation}
x_{a}(\ubr)=\xbr_{a}\,,\quad \xi\equiv\xi_{\L}=\frac{1}{\xi_{\R}}= {\ka\ov2\pi}+ \sqrt{1+{\frac{\ka^2}{4\pi^2}}}
\,.
\end{equation}
The  branch point $\ubr$ is of the square-root type, and  going around it,  $x_\L(u)$ and $x_\R(u)$  invert as
\begin{equation}
\label{eq:crossingpath}
x_a^\circlearrowleft(u)=\frac{1}{x_{\bar a}(u)}\,,
\end{equation}
where $x^\circlearrowleft$ indicates analytic continuation along a path around the  branch point $\ubr$. Note that then $x_{a}^\pm\to 1/x_{\bar{a}}^\pm$, $p_{a}\to -p_{\bar{a}}$, and $E_{a}\to -E_{\bar{a}}$, as expected for the crossing transformation~\cite{Lloyd:2014bsa}. 

\begin{figure}[t]
\includegraphics[width=4.2cm]{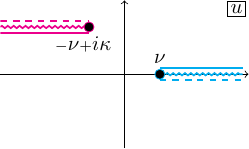}%
\hspace{2mm}%
\includegraphics[width=4.2cm]{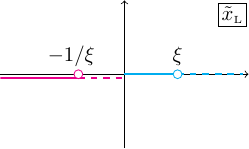}\\
\includegraphics[width=4.2cm]{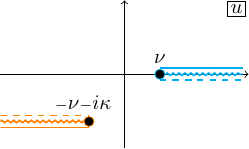}%
\hspace{2mm}%
\includegraphics[width=4.2cm]{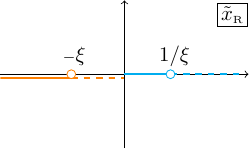}\\%
\caption{\label{fig:mirror}%
Analytic structure of $\tilde{x}_{\L,\R}(u)$. Left top/bottom: each of $\tx_{\L}(u)$ and $\tx_{\R}(u)$ has two branch cuts; exchanging $L\leftrightarrow R$ is a reflection about the real-$u$ line. Right top/bottom: the images of the $u$-plane cuts are straight lines on the $\tilde{x}_{\L,\R}$ plane; For $\tilde{x}<0$ they run just below the cut of~$\ln\tilde{x}$.
Sending $\kappa\to0$ one immediately recovers the familiar structure of the pure-RR mirror theory~\cite{Frolov:2021bwp}.
The lower-half $\tx_{\L,\R}$-plane is the physical region for mirror particles (``mirror'' region), and the upper half-plane for antiparticles (``antimirror'' region).}
\end{figure}

\paragraph{``Mirror'' kinematics.}
To describe the  spectrum  of this model by integrability we need to introduce an auxiliary ``mirror'' model~\cite{Arutyunov:2007tc}. Kinematically, it is obtained by a double Wick rotation
\begin{eqnarray}
   {1\ov i} p_a\to \tilde{E}_a\,,\qquad
   {1\ov i} E_a\to \tilde{p}_a\,,
\end{eqnarray}
This transformation makes the dispersion~\eqref{eq:dispersion} complex which, while unusual, is not problematic~\cite{Baglioni:2023zsf}.
We demand that the ``mirror'' S~matrix, including its dressing factors, can be obtained from the original (``string'') one by analytic continuation, as it is the case at $k=0$~\cite{Arutyunov:2009kf,Frolov:2021fmj}.
This can be done by introducing mirror variables
\begin{equation}
\tx_a(u)=
\begin{cases}
x_a(u)&\text{Im}[u]<0\,,\\
1/x_{\bar{a}}(u)&\text{Im}[u]>0\,,
\end{cases}
\end{equation}
and continuing $x_a\to \tx_a$. Note that $\tx_a(u)^*={1/ \tx_{\bar{a}}(u^*)}$, differently from the string kinematics.
Perhaps surprisingly, the mirror kinematics is simpler than the string one: on the $u$-plane we have two branch cuts for $\tx_{\L}(u)$ and  $\tx_{\R}(u)$, while on the $\tx_{\L,\R}$ plane the physical region is simply the lower half-plane, see Figure~\ref{fig:mirror}.
Hence, we will construct the dressing factors \textit{in the mirror region first} and then continue them to the string region.

\paragraph{Crossing equations.}
The crossing equations for S-matrix elements were derived in \cite{Lloyd:2014bsa}.  Their normalisation however introduces unphysical zeros and poles in the scattering matrix (see also~\cite{OhlssonSax:2023qrk}); Instead we set, following the notation of~\cite{Lloyd:2014bsa},
\begin{eqnarray}
\label{eq:massivenorm}
    \displaystyle
    \mathbf{S}|\bar{Z}_{1}\bar{Z}_{2}\rangle
    \displaystyle&=&
    \frac{u_{12} + {2i\ov h}}{u_{12} - {2i\ov h}}
  e^{-2i\tilde{\phase}^{11}_{\R\R}}\,
    \big|\bar{Z}_{1}\bar{Z}_{2}\big\rangle,\\
\nonumber
\displaystyle
    \mathbf{S}|\bar{Z}_{1}Y_{2}\rangle\displaystyle&=&
     {x_{\R1}^-\ov x_{\R1}^+}
    \frac{1-x_{\R1}^+x_{\L2}^+}{1-x_{\R1}^-x_{\L2}^-}
    \frac{1-x_{\R1}^+x_{\L2}^-}{1-x_{\R1}^-x_{\L2}^+}e^{-2i\tilde{\phase}^{11}_{\R\L}}
    |\bar{Z}_{1}Y_{2}\rangle.
\end{eqnarray}
where  $u_{12}\equiv u_1-u_2$, 
$x_{aj}^\pm = x_a(u_j \pm {i\ov h})$, $a=$L,R, $j=1,2$. Here $\tilde{\phase}^{11}_{ab}$ is the dressing phase. The normalisation for the scattering $Z_1$-$Z_2$ and $Z_1$-$\bar{Y}_2$ is obtained from the one above by exchanging L and R labels. From it we can also read the normalisation for $Y_1$-$Y_2$ and $Y_1$-$\bar{Z}_2$ since the scattered states belong to the same representations of the supersymmetry algebra.
In this normalisation we find that 
the dressing factors satisfy  the following  crossing relations
\begin{equation}\label{eq:crossing}
\begin{aligned}
&\exp\left[2i\tilde{\phase}^{11}_{aa} \left(x_{a1}^\pm,x_{a2}^\pm\right)+2i\tilde{\phase}^{11}_{\bar aa}\left(1/x_{a1}^{\pm},x_{a2}^\pm\right)\right]
\\
&\qquad\qquad=
\frac{(x_{a1}^- - x_{a2}^+)(x_{a1}^+ - x_{a2}^-)}{(x_{a1}^- - x_{a2}^-)(x_{a1}^+ - x_{a2}^+)}
  \frac{u_{12} + {2i\ov h}}{u_{12} - {2i\ov h}}
\,,\\
&\exp\left[2i\tilde{\phase}^{11}_{\bar{a}a} \left(x_{\bar{a}1}^\pm,x_{a2}^\pm\right)+2i\tilde{\phase}^{11}_{aa}\left(1/x_{\bar{a}1}^{\pm},x_{a2}^\pm\right)\right]
\\\
&\qquad\qquad=
\frac{\big(1-x^+_{\bar a1}x^+_{a2}\big)\big(1-x^-_{\bar a1}x^-_{a2}\big)}{\big(1-x^+_{\bar a1}x^-_{a2}\big)\big(1-x^-_{\bar a1}x^+_{a2}\big)}
 \frac{u_{12} + {2i\ov h}}{u_{12} - {2i\ov h}}
 \,,
\end{aligned}
\end{equation}
where the points $1/x^{\pm}_1$ in the phases are reached by analytically continuing $\tilde{\theta}(x^{\pm}_1, x^{\pm}_2)$ to the antistring region, cf.\ Figure~\ref{fig:string}.

\paragraph{$\ka$-deformed phases.}
\label{sec:crkaBES}
Here we propose $\ka$-deformed Beisert-Eden-Staudacher (BES) \cite{Beisert:2006ez} phases  and the corresponding $\ka$-deformed Hern{\'a}ndez-L{\'o}pez (HL) \cite{Hernandez:2006tk} phases; Specifically, we consider the ``improved'' BES phase, in the sense of~\cite{Arutyunov:2009kf}. Starting from the mirror kinematics we will construct~$\tilde{\theta}^{11}_{ab}(\tx_1^\pm,\tx_2^\pm)$, which may then be continued to the string region to~$\tilde{\theta}^{11}_{ab}(x_1^\pm,x_2^\pm)$ appearing above.
To begin with, we assume that all such $\tilde{\theta}(\tx_{1}^\pm,\tx_{2}^\pm)$ may be decomposed as~\cite{Arutyunov:2006iu}
\begin{eqnarray}
\label{eq:decomposition}
    \tPhi(\tx_{1}^+,\tx_{2}^+)-\tPhi(\tx_{1}^+,\tx_{2}^-)
    -\tPhi(\tx_{1}^-,\tx_{2}^+)+\tPhi(\tx_{1}^-,\tx_{2}^-)\,,
\end{eqnarray}
for some suitable function~$\tPhi(\tx_{1},\tx_{2})$. This ensures good properties under bound-state fusion.
Next, we define the functions~$\tPhi(\tx_1,\tx_2)$ by an integral representation reminiscent of~\cite{Dorey:2007xn}, adapted to our mirror model:
\begin{equation}\begin{aligned}
\label{PhiBESka}
&\tPhi_{ab}^{\alpha\beta}(\tx_1,\tx_2)
\\
&\quad=-\int\displaylimits_{\pa\cR_{\alpha}} \frac{{\rm d}w_1}{2\pi i}\int\displaylimits_{\pa\cR_{\beta}}\frac{{\rm
d}w_2}{2\pi i}\frac{K(u_a(w_1)-u_b(w_2))}{(w_1-\tx_1)(w_2-\tx_2)}\,,
\end{aligned}\end{equation}
where $K(v)$ is a suitable Kernel and $\partial\mathcal{R}_\alpha$ denotes the boundary of the mirror physical region or of its complement (the lower/upper half-plane). Because $u_a(w)$ features the cut of $\ln w$,  $\partial\mathcal{R}_\alpha$ can run just above the cut ($\alpha=+$), or just below ($\alpha=-$).
For the singularities at $w=0$ and $w=\infty$ we adopt the principal value prescription, which leads to convergent integrals when considering the combination~\eqref{eq:decomposition}.
To obtain the BES phase we take $K(v)$ to be the BES Kernel,
\begin{equation}\la{regBESkernel}
K^{\bes}(v)\equiv i\,\log\frac{\Gamma\big(1+\tfrac{ih}{
2}v\big)}{
\Gamma\big(1-\tfrac{ih}{
2}v\big)}\,,
\end{equation}
and use resulting integrals to define
\begin{equation}
\label{eq:semisum}
\begin{aligned}
\tPhi_{aa}^{\bes}(\tx_1,\tx_2)\equiv&\frac{\tPhi_{aa}^{++,\bes}(\tx_1,\tx_2)+\tPhi_{aa}^{--,\bes}(\tx_1,\tx_2)}{2},\\
\tPhi_{a\bar{a}}^{\bes}(\tx_1,\tx_2)\equiv&\frac{\tPhi_{a\bar{a}}^{+-,\bes}(\tx_1,\tx_2)+\tPhi_{a\bar{a}}^{-+,\bes}(\tx_1,\tx_2)}{2}.
\end{aligned}
\end{equation}
The standard HL Kernel for the double-integral representation is singular~\cite{Borsato:2013hoa} but we may regularise it as
\begin{equation}
    K^{\hl}(v)= \frac{\ln(\epsilon-iv)-\ln(\epsilon+iv)}{2i}\,,
\end{equation}
and take $\epsilon\to0$ after integration. We then define $\tPhi_{aa}^{\hl}(\tx_1,\tx_2)$ as in~\eqref{eq:semisum}, though the two terms in each semi-sum happen to be equal in the HL case. 

\paragraph{Difference-form phases.}
Like in the pure-RR case~\cite{Frolov:2021fmj}, we expect part of the dressing factors to be of difference form. We introduce a function $R(\gamma)$ defined in terms of Barnes $G$-functions,
\begin{equation}
\label{eq:Barnes}
R (\g)= {G(1- \frac{\g}{2\pi i})\ov G(1+ \frac{\g}{2\pi i}) } \,,
\end{equation}
which satisfies simple monodromy relations under~$\gamma\to\gamma\pm2\pi i$~\cite{Frolov:2023lwd}.
Then,  the $\g$-rapidities generalising the ones used in \cite{Beisert:2006ib,Fontanella:2019baq,Frolov:2021fmj} are expressed in terms of the Zhukovsky variables as
\begin{equation}
\g_a(x)=\ln\frac{x-\xi_a}{x\xi_a+1}\,,\quad
\tg_a(\tx)=\ln\frac{\xi_a-\tx}{\tx\xi_a+1}-\frac{i\pi}{2}\,.
\end{equation}
This follows by requiring the branch points of $\gamma_a(x)$ to correspond with those of $x_a(u)$, cf.~\eqref{uplane},  and that under the crossing map~\eqref{eq:crossingpath} $\g$ and $\tg$ transform as $\g_a\to \g_{\bar{a}}- i\pi$ and  $\tg_a\to \tg_{\bar{a}}- i\pi$.
Continuing $\gamma_a^\pm$ from the string to the mirror region yields a shift of~$-i\tfrac{\pi}{2}$. Hence, difference-form expressions are unchanged, which makes their analytic continuation straightforward.
Introducing the short-hand $\tg_{ab} \equiv \tg_a(\tx_{a1}) -\tg_b(\tx_{b1})$ with $a,b=\text{L,R}$, we define the ``odd phase'' (in the sense of~\cite{Beisert:2006ib}) as
\begin{equation}
\begin{aligned}
&\Phi^{\barnes}_{aa}(\tx_{a1},\tx_{a2})=+i\ln R(\gamma_{aa})\,,\\
&\Phi^{\barnes}_{a\bar{a}}(\tx_{a1},\tx_{\bar{a}2})=-\frac{i}{2}\ln R(\gamma_{a\bar{a}}+i\pi)R(\gamma_{a\bar{a}}-i\pi)\,,
\end{aligned}
\end{equation}
with~$\tilde{\theta}^{\barnes}_{ab}(\tx_{a1}^\pm,\tx_{b2}^\pm)$ given by~\eqref{eq:decomposition}. 

\paragraph{Proposal for the dressing factors.}
With these ingredients, we define the dressing factors of $m=1$ mirror particles as
\begin{equation}
\label{eq:solution}
\begin{aligned}
\tilde{\theta}^{11}_{ab}(\tx_{a1}^\pm,\tx_{b2}^\pm)
=&+\tilde{\theta}^{\bes}_{ab}(\tx_{a1}^\pm,\tx_{b2}^\pm)-\tilde{\theta}^\hl_{ab}(\tx_{a1}^\pm,\tx_{b2}^\pm)\\
&+\tilde{\theta}^{\barnes}_{ab}(\tx_{a1}^\pm,\tx_{b2}^\pm)\,.
\end{aligned}
\end{equation}
The formula is given for fundamental particles but, owing to~\eqref{eq:decomposition}, it can be  extended to the scattering of $ (m_1,m_2)$-particle bound states, at least when $m_j=|M_j|\neq0$ mod$k$, as that case has a different (massless) kinematics, cf.~\eqref{eq:dispersion}.

\paragraph{Properties.}
We checked that~\eqref{eq:solution} solves the mirror version of~\eqref{eq:crossing} (which just amounts to continuing $x\to\tx$ in the right-hand side of that equation). 
We also computed the ``string'' dressing factor by analytic continuation of~\eqref{eq:solution} and checked that it satisfies~\eqref{eq:crossing}; The analytic continuation is conceptually straightforward, though nontrivial. Our proposal reduces to the ones of~\cite{Frolov:2021fmj} when~$\kappa\to0$. Its ``odd'' part $\theta_{ab}^{\barnes}(x_{a1}^\pm,x_{b2}^\pm)$ agrees with the proposal of~\cite{OhlssonSax:2023qrk}.  
Crossing holds for bound states too, both in the string and mirror models. Left-right symmetry~\cite{Lloyd:2014bsa} and braiding unitarity hold by construction, and physical unitarity also holds in the string kinematics.
It is also easy to see that the mirror dressing factors~\eqref{eq:solution} are invariant under P~(parity). In the string model, P-invariance is broken when $k\neq0$ but CP holds (C is charge conjugation). This is subtle because the string physical region of Figure~\ref{fig:string} has $0\leq p_a<2\pi$. To verify CP one must suitably continue the dressing factors to $p_a<0$. This can be done and CP indeed holds. Finally, note that~\eqref{eq:dispersion} is invariant under~$M\to M\pm k$ and $p\to p\mp 2\pi$. We have found that in the string kinematics the full S~matrix, including~\eqref{eq:solution}, is identical when swapping a $m$-particle bound state of momentum~$p$ for an $(m+k)$-particle bound state of momentum $p\pm 2\pi$, where the plus sign is for R-particles and the minus for L-particles. Hence, such particles are indistinguishable.
This type of periodicity is a novelty of the mixed-flux model, and it is due to the linear term in~\eqref{eq:dispersion}. It allows to equivalently obtain bound states from analytic continuation in $p$, as well as by the ``fusion'' procedure~\cite{Frolov:2025uwz}. By contrast, in pure-RR models such as $AdS_5 \times S^5$ the periodicity is just $p\pm 2\pi$ and $M\to M$ so that bound states can \textit{only} be obtained by fusion~\cite{Arutyunov:2009kf}.
Finally, remark that our proposal is a ``minimal'' solution of the crossing equations; As always, more solutions to the crossing equations can be found by means of CDD factors~\cite{Castillejo:1955ed}. However, we could not find any such factors which would not spoil the symmetries, analytic structure, and perturbative behavior of the S~matrix, as we discuss in~\cite{Frolov:2025uwz}.

\paragraph{Large-tension expansion and comparison with the string sigma model.}
Here we compare the S~matrix given by~\eqref{eq:solution}  with the perturbative worldsheet S~matrix at tree-level (mirror and string) and one-loop (string).  
Expanding the BES kernel at large $h$ keeping $\ka$ fixed, one gets
\begin{equation}
\label{eq:BESexpanded}
K(v)= h\, K^\afs(v) + K^\hl(v)+\cO(h^{-1})\,,
\end{equation}
where
\begin{equation}
\label{eq:AFSkernel}
K^\afs(v) =\frac{(\epsilon-iv)\ln(\epsilon-iv)-(\epsilon+iv)\ln(\epsilon+iv)}{2i},
\end{equation}
and we have dropped from $K^\afs$ terms which do not contribute to the $\tilde{\theta}^{\afs}$; Note that the HL order precisely cancels the term from $\tPhi^{\hl}$ in~\eqref{eq:solution}.
The calculation of the AFS phases is straightforward though lengthy. In the mirror region we get
\begin{eqnarray}
\tPhi_{aa}^{\afs}(\tx_1,\tx_2)
=-\frac{\kappa_a}{4 \pi }\text{Li}_2\frac{\tx_1-\tx_2}{\tx_1} +\frac{1}{2}\frac{\ln \tx_2}{\tx_1}
\\
\nn
+\frac{1}{2} u_{1} \ln \frac{u_{1}-u_{2}}{\tx_1-\tx_2}   -   (1\leftrightarrow 2)
 \,,
\end{eqnarray}
and
\begin{eqnarray}
\tPhi_{a\bar a}^{\afs}(\tx_1,\tx_2)
= \frac{\kappa_a}{4\pi}\left[\ln \tx_1\ln \tx_2+2\text{Li}_2\frac{\tx_1\tx_2-1}{\tx_1\tx_2}\right]\quad\\
\nn
+\frac{1}{2}
   \left[\frac{\ln \tx_2}{\tx_1}+\frac{\tx_1{}^2+1}{\tx_1}\ln\frac{\tx_1\tx_2-1}{\tx_1\tx_2}-(1\leftrightarrow2)\right].
\end{eqnarray}
Note that $\tx_1$, $\tx_2$ can be trivially taken to the string region, at least for $\text{Re}[x_j]>0$, \textit{i.e.} for $\text{Re}[p_j]<\pi$.  
%
%
%
Before comparing our AFS phases to those obtained by using the finite-gap equations \cite{Babichenko:2014yaa} and the classical scattering matrix of bound states \cite{Stepanchuk:2014kza}, it is necessary to account for the normalisation of the S~matrix. Then, up to a choice of branches of logarithms, $\Phi^{\afs}_{ab}$ agree with the corrected phases from \cite{Babichenko:2014yaa,Stepanchuk:2014kza}.

\paragraph{Near-BMN expansion.}
To compute the near-BMN expansion we take the tension $T\gg1$, with
\begin{equation}
\label{eq:tension}
k =2\pi q \, T\, , \qquad h= \sqrt{1 - q^2}\,T+\cO(T^{-1}) \,,
\end{equation}
and the momenta small, of order $\mathcal{O}(T^{-1})$. The terms of order $1/T$ in \eqref{eq:tension} contribute to the S~matrix from two loops.
At tree level, our results match with the perturbative ones~\cite{Hoare:2013pma,Engelund:2013fja,Hoare:2013lja,Roiban:2014cia,Babichenko:2014yaa,Bianchi:2014rfa,Sundin:2014ema,Baglioni:2023zsf} in the string and mirror model. This is true both for small and positive, as well as small and negative, momentum (the latter requires an analytic continuation to reach $p_a<0$).
At one-loop we find a mismatch for the difference of S-matrix elements of the form
\begin{equation}
\delta\Phi_{ab}=\pm
\frac{i}{4\pi T^2} p_ap_b\,(p_a\omega_b-p_b \omega_a)+\mathcal{O}(T^{-3})\,,
\end{equation}
where $\omega_a(p)=\sqrt{p^2+2M_a qp+1}$ is the dispersion relation of the particles in the limit $T\to\infty$. This is the same mismatch as in the pure RR case \cite{Frolov:2021fmj} and it can be removed by the same local counterterm as in that case, owing to the fact that there is no dependence on the fluxes other than through~$\omega$.
\paragraph{Relativistic limit.}
This model has a nice relativistic limit~\cite{Frolov:2023lwd} if we expand around the minimum of the dispersion relation,
\begin{equation}
p= -\frac{2\pi  M}{k}+h\,\delta p+\mathcal{O}(h^2)\,.
\end{equation}
At $\mathcal{O}(h)$ the model is relativistic, with massive and massless excitations depending on whether $M\neq0$ mod$k$. The S~matrix and dressing factors can be bootstrapped~\cite{Frolov:2023lwd} as in any relativistic integrable QFT. Interestingly, this limit involves \textit{small} rather than large~$h$, as well as a different notion of crossing than~\eqref{eq:crossingpath}, as also discussed in~\cite{OhlssonSax:2023qrk}. Nonetheless, the expansion of~\eqref{eq:solution} matches the dressing factors of~\cite{Frolov:2023lwd}.

\paragraph{Outlook.}
\label{conclusions}
We proposed a solution of the mixed-flux crossing equations for massive particles which is consistent with the symmetries of the model and passes several checks. These involve a number of explicit computations which we shall present elsewhere~\cite{Frolov:2025uwz}.
Our proposal automatically yields bound-state dressing factors in the mirror and string model. Because states with $m=0$ mod$k$ have a gapless dispersion, they can be studied to get insight into the dynamics of massless particles, which is necessary to solve the massless crossing equations~\cite{Lloyd:2014bsa}. Moreover, for massive excitations we found that a $(m+k)$-particle bound-state is equivalent to a $m$-particle bound-state at shifted momentum. If this also holds for $m=0$ mod$k$, it will be straightforward to obtain the massless dressing factors from fusion, counterintuitive as this may seem.
The dressing factors are the missing ingredient needed to obtain the spectrum of this model. It will be especially interesting to study them at $k$ fixed and $h\ll1$, which amounts to perturbing around pure-NSNS backgrounds~\cite{Maldacena:2000hw,Dei:2018mfl}, see also~\cite{Cho:2018nfn}. The case $k=1$ and $h\ll1$, in particular, will allow us to make contact with the perturbations around the symmetric-product orbifold CFT, see~\cite{Seibold:2024qkh} for a recent review.

\begin{acknowledgments}
We thank the participants of the Workshop ``Integrability in Low Supersymmetry Theories'' in Filicudi, Italy for stimulating discussions that initiated this project. D.P. thanks Bogdan Stefański for stimulating discussions during the 2024 workshop on `Gauge theories, supergravity and superstrings' in Benasque.
D.P.~and A.S.~acknowledge support from the EU - NextGenerationEU, program STARS@UNIPD, under project ``Exact-Holography'', and from the PRIN Project n.~2022ABPBEY. A.S.~also acknowledges support from the CARIPLO Foundation under grant n.~2022-1886, and from the CARIPARO Foundation Grant under grant n.~68079.
A.S.~also thanks the MATRIX Institute in Creswick \& Melbourne, Australia, for support through a MATRIX Simons fellowship in conjunction with the  program ``New Deformations of Quantum Field and Gravity Theories'', as well as the IAS in Princeton for hospitality. This work has received funding from the Deutsche Forschungsgemeinschaft (DFG, German Research Foundation) -- SFB-Geschäftszeichen 1624 – Projektnummer 506632645. 
S.F.~acknowledges support from the INFN under a Foreign Visiting Fellowship at the last stages of the project.
\end{acknowledgments}


\bibliographystyle{apsrev4-1}

\bibliography{refs}

\end{document}